\documentclass[preprint,11pt,authoryear,times,nopreprintline]{elsarticle}

\usepackage[T1]{fontenc}
\usepackage[utf8]{inputenc}
\usepackage{microtype}
\usepackage{graphicx}
\usepackage{geometry}
\usepackage{amssymb}
\usepackage{amsmath}
\usepackage{caption}
\biboptions{authoryear}

\usepackage[figuresright]{rotating}

\usepackage[bookmarks=false]{hyperref}
\hypersetup{colorlinks,
  linkcolor=blue,
  citecolor=blue,
  urlcolor=blue}

\journal{Transportation Research Procedia}

\begin{document}
\begin{frontmatter}

\title{Agent-Based Transport Model Calibration via Online Stochastic Batch Optimization\tnoteref{t1}}
\tnotetext[t1]{Presented at the World Conference on Transport Research (WCTR 2026), Toulouse, France.}

\author[a]{Abdelkader Dib\corref{cor1}}
\ead{dabdelkader@ethz.ch}
\author[a]{Milos Balac}

\address[a]{Center for Sustainable Future Mobility (CSFM), ETH Zürich, Switzerland}

\cortext[cor1]{Corresponding author. Tel.: +41 78 446 11 95.}

\begin{abstract}
Calibrating agent-based models (ABMs) is computationally demanding, as traditional black-box optimization typically requires hundreds or thousands of independent simulation runs. To reduce this burden, we introduce an online calibration framework that estimates parameters within a single simulation run. The approach combines black-box optimization with stochastic batch optimization, a method widely applied in machine learning but not yet explored in ABM calibration. Using MATSim as a case study, we treat each simulation iteration as a stochastic batch, enabling incremental parameter updates as agents make decisions. This online calibration paradigm allows model parameters and agent behaviors to co-evolve, achieving convergence without restarting the simulation. We demonstrate the effectiveness of our method by calibrating utility function parameters in MATSim for a large-scale transportation scenario of Switzerland. Results show that mode shares align closely with observed data within a typical simulation timeframe, confirming the feasibility and efficiency of the approach. By embedding calibration directly into the simulation loop, our method opens new possibilities for real-time, scalable, and efficient calibration of complex ABMs.
\end{abstract}

\end{frontmatter}

%\correspondingauthor[*]{Corresponding author. Tel.: +0-000-000-0000 ; fax: +0-000-000-0000.}

% \linenumbers

%\vspace*{-12pt}
%\enlargethispage{-7mm}

\section{Introduction}
Agent-based modeling (ABM) has become a key tool in transportation planning for simulating the unique characteristics and complex behaviors of individual travelers and their interactions \citep{bonabeau2002, reviewAgentBasedTransportModels}. By representing each traveler as an agent with detailed attributes, plans, and decision rules, ABMs can capture heterogeneous responses to policy interventions, network changes, and emerging mobility solutions \citep{amodParis, becker2020assessing}.

A major challenge for the practical deployment of ABMs is calibration \citep{advancingCalibration, reviewAgentBasedTransportModels, empiricalValidation}, the process of adjusting model parameters so that simulated outputs align with real-world observations. Calibration involves running the model with varying parameters and evaluating performance by comparing outputs against observed data, with the goal of finding parameter sets that minimize discrepancies between simulation and reality \citep{empiricalValidation}.

For instance, in MATSim (Multi-Agent Transport Simulation \citep{matsim}), an open-source, agent- and activity-based modeling framework that simulates individual travel behavior within transportation networks, mode choice models rely on utility functions with multiple parameters such as alternative-specific constants or marginal utilities. While these parameters can be initially estimated from survey data, they often require adjustment to align simulated mode shares with observed values. Such adjustments compensate for (i) model discrepancy, as models inevitably simplify real-world systems and cannot perfectly reproduce them \citep{CalUncertainty}, and (ii) potential biases or limitations in survey data, which may be incomplete or unrepresentative, necessitating the inclusion of additional data sources. Furthermore, unmeasured factors not explicitly modeled in MATSim \citep{berlin}, such as comfort, public transport crowding, or perceived safety, can strongly influence travel behavior and vary across different regions. Consequently, when extracting a regional model from a larger model, calibration is necessary to account for these regional differences.

The standard calibration workflow for large-scale ABMs such as MATSim is computationally intensive. Each candidate parameter set requires a full simulation run to generate output metrics (e.g., mode shares, traffic counts) for comparison with observed data. A new parameter set is then proposed, and the iterative process continues until the output metrics fall within an acceptable range. Calibration methods typically rely either on manual parameter tuning by domain experts \citep{eqasim, berlin0, spgm, melbourne} or on “black-box” optimization algorithms that explore the parameter space without explicit knowledge of the model structure \citep{horl2022exploring, liangcalibration}. Although these methods can be effective, they often require hundreds or thousands of full simulation runs. For large-scale applications, where simulations may involve millions of agents and complex networks, with a single run lasting several hours or more, this approach becomes impractical.

This paper introduces an alternative paradigm: online calibration for large-scale ABMs using stochastic batch optimization, a technique inspired by the machine learning (ML) field but, to the best of our knowledge, not previously applied to ABM calibration. Rather than executing multiple independent simulations within a black-box optimization loop, the calibration process is embedded directly into the simulation’s iterative dynamics. In this way, parameter adjustment occurs within a single run.

In this work, we propose a single-run, iterative calibration framework that shifts the focus from multi-run black-box optimization to online stochastic batch optimization.

The remainder of this paper is structured as follows: \autoref{sec:relatedWork} reviews related work on ABM calibration and estimation methods, \autoref{sec:background} introduces eqasim \cite{eqasim} and MATSim, which are used in our use cases, and outlines the concept of stochastic batch optimization, \autoref{sec:methodology} formalizes the proposed calibration algorithm, \autoref{sec:experiments} reports experimental results on a large-scale transportation scenario, and \autoref{sec:discussion} provides a discussion of the findings. Finally, \autoref{sec:conclusion} concludes with implications and directions for future research.

\section{Related Work}
\label{sec:relatedWork}
The use of ABMs is widespread across disciplines, including economics \citep{economic1, economic2}, sociology \citep{sociology1}, biology \citep{biology1, biology2}, ecology \citep{ecology1}, epidemiology \citep{advancingCalibration}, energy systems \citep{energySystems1}, land-use \citep{landUse}, emission modeling \citep{tang2015carbon,dib2023two,spgem}, and transportation \citep{reviewAgentBasedTransportModels, reviewAbmTransportation}. They are valued for their ability to capture complex system dynamics and generate predictive insights.

Despite their abstraction of reality through simplified variables and interactions \citep{empiricalValidation}, ABMs often involve a large number of parameters that must be calibrated to ensure alignment between simulated outputs and observed data. Calibration is typically framed as a parameter search problem, where the objective is to minimize the discrepancy between model outputs and empirical observations. However, the stochastic and computationally intensive nature of ABMs makes this process challenging. Early approaches relied on trial-and-error, grid search, or random sampling, but these methods become impractical as model complexity grows \citep{advancingCalibration}. Consequently, more advanced optimization techniques are required.

To automate calibration processes, researchers have increasingly applied black-box optimization algorithms. Among these, a widely used class are evolutionary and population-based search methods \citep{horl2022exploring}. These approaches maintain a population of candidate solutions that are iteratively updated. In each generation, new candidates are generated through variation operators such as mutation, recombination, or sampling from adaptive distributions, with selection mechanisms favoring parameters that performed well in earlier iterations. Representative algorithms include genetic algorithms \citep{genericAlgo1, genericAlgo2}, particle swarm optimization (PSO) \citep{pso1}, and Covariance Matrix Adaptation Evolution Strategy (CMA-ES) \citep{horl2022exploring}.

Other optimization strategies have also been explored. For instance, Bayesian optimization methods \citep{bayesian1, bayesian2} balance exploration and exploitation using probabilistic surrogate models, while Simultaneous Perturbation Stochastic Approximation (SPSA) \citep{liangcalibration, spsa1} provides a gradient-free stochastic approach particularly suitable for high-dimensional problems.

A complementary line of research focuses on surrogate or emulator models to mitigate the computational burden of repeated simulation runs. These approaches construct fast approximation models that predict ABM outputs as functions of input parameters \citep{surrogate1, surrogate2, surrogate3}. Prominent surrogates include Gaussian Processes (GPs), support vector machines (SVMs), random forests, and deep neural networks. While surrogate-based approaches have demonstrated considerable success, their effectiveness depends on careful surrogate selection, optimization, and validation, particularly in high-dimensional settings where complex models such as neural networks are prone to overfitting.

In the transportation context, several specialized calibration frameworks have been proposed, although they remain uncommon in standard practice. \cite{liangcalibration} treated MATSim as a black-box model and applied SPSA to estimate utility parameters. Similarly, \cite{horl2022exploring} introduced a general MATSim calibration framework, implementing various optimization algorithms including CMA-ES, SPSA, and Natural Evolution Strategies (NES). \cite{agarwal2017calibration} presented a proof of concept for calibrating MATSim mode choice parameters, employing a search-acceleration strategy originally proposed by \cite{flotterod2017search}. \cite{efficientCalibrationFlotterod,agriesti2023bayesian} used surrogate models to accelerate calibration and reduce the number of simulation runs. A more targeted approach was implemented by \cite{berlin}, which applied a formula from \cite{berlinFormula} designed specifically for calibrating alternative-specific constants (ASCs) for MATSim. In this iterative approach, new ASCs are computed after each full MATSim run based on the simulated mode shares, followed by a simulation rerun with the updated ASCs. The procedure is repeated until convergence is achieved. In their case, a stable solution was reached after only a few iterations.

Our proposed method differs from the existing frameworks by abandoning multi-run approaches entirely. Instead of conducting repeated independent simulations, we adopt an online calibration approach, where parameters are continuously updated during the execution of the simulation. To our knowledge, this strategy has not previously been applied in this context. 

The next section provides the background to our methodology, which is then presented in \autoref{sec:methodology}.

\section{Background} 
\label{sec:background}

This section introduces the key components relevant to our study. We begin with the eqasim pipeline, which is used to generate MATSim scenarios. We then describe MATSim and its discrete mode choice extension, which together form the foundation of our use case studies. Finally, we outline the concept of stochastic batch optimization, a technique widely used in ML, which serves as the methodological inspiration for our approach.

\subsection{eqasim}

The eqasim pipeline \citep{eqasim}, is a modular, open-source framework for generating synthetic populations and their travel demand from publicly available data. It combines data sources such as census records, travel surveys, and OpenStreetMap to create a population that reflects the demographic and spatial characteristics of real-world populations, and the transportation infrastructure. Each agent is described by detailed attributes, including age, income, sex, ownership of a driving license, and public transport subscription. In addition, the framework generates a daily activity plan for each agent. These plans consist of sequences of activities located in time and space, linked by corresponding trips.

A key feature of eqasim is its ability to convert the generated population and activity plans into MATSim input formats, enabling large-scale transport simulations for the corresponding region. 

The platform provides ready-to-use scenarios at both urban and national scales, including Île-de-France (France) \citep{idfEqasim}, São Paulo (Brazil) \citep{saoPauloEqasim}, California (USA) \citep{californiaEqasim}, and Switzerland \citep{switzerlandScenario}.

\subsection{MATSim}
MATSim (Multi-Agent Transport Simulation) is an open-source, agent-based simulation framework designed to model large and complex transportation systems \citep{matsim}. The model requires two primary inputs: a representation of the transportation infrastructure (including road and transit networks, schedules, and capacities) and a synthetic population, where each agent represents an individual with specific attributes and a detailed plan of daily activities. These activities are connected by trips that represent the travel demand within the network.

MATSim operates through a co-evolutionary, iterative process that simulates daily mobility patterns and gradually optimizes agent travel plans until reaching equilibrium. In each iteration, a subset of agents are selected to replan their daily activities and travel choices. Several replanning strategies exist, including adjusting departure times, selecting alternative routes, or switching transport modes, allowing agents to randomly explore many options that could improve their daily plans. The updated plans are then evaluated in subsequent simulation runs, creating a feedback loop that drives the system toward equilibrium, where system variables stabilize within acceptable ranges across iterations.

\subsection{Discrete Mode Choice Modeling in MATSim}
\label{sec:dmc}

\begin{figure}[h]
    \centering
    \includegraphics[width=0.6\linewidth]{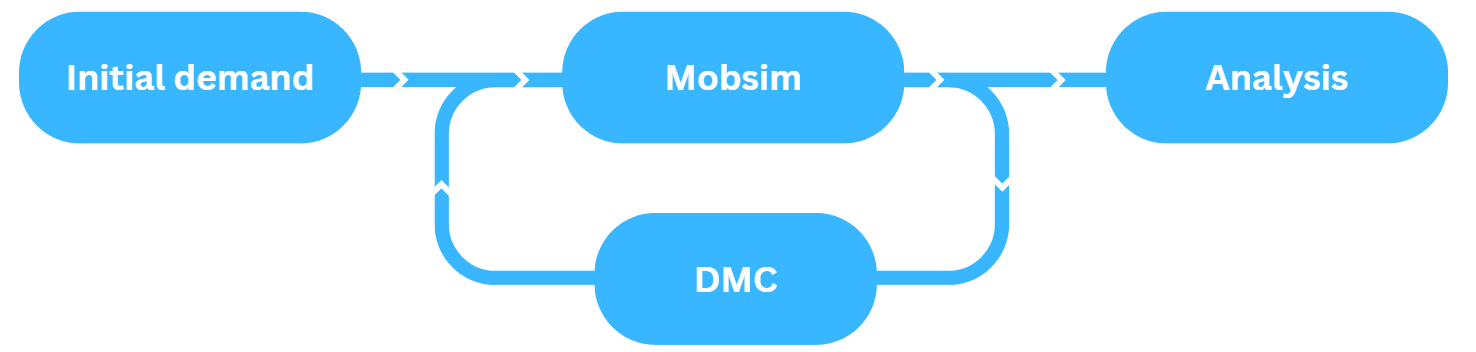}
    \caption{Discrete Mode Choice in MATSim}
    \label{fig:dmc}
\end{figure}

A key component of MATSim is the discrete mode choice (DMC) extension \citep{dmc1,dmc2}, which allows agents to dynamically adapt their travel behavior across simulation iterations using a discrete choice model. As illustrated in \autoref{fig:dmc}, during each iteration, a subset of agents is selected to re-evaluate their transport mode choice from a set of alternatives, such as car, walking, cycling, and public transport. This decision is based on specific choice variables, including travel times and other relevant attributes. Unlike the standard random replanning process, the DMC extension enables informed replanning, preventing agents from selecting unrealistic options and thereby accelerating convergence. Through this iterative process, agents continuously adjust their choices in response to evolving network conditions, ultimately leading the system toward an equilibrium.

Each transport mode is characterized by a vector of choice variables (e.g., travel time, access egress time, monetary cost), which collectively influence the agent's decision-making process. The selection mechanism is formalized using discrete choice models with the Multinomial Logit (MNL) model. In the MNL framework, the total utility $ U_{nj} $ that agent $ n $ derives from choosing mode alternative $ j $ is modeled as a random variable:
\begin{equation}
    U_{nj} = V_{nj} + \varepsilon_{nj},
    \label{eq:utility}
\end{equation}
where $ V_{nj} $ represents the deterministic component of utility, reflecting observable factors, and $ \varepsilon_{nj} $ denotes the stochastic error term, capturing unobserved influences on preferences and perception errors. It is assumed that the error terms $ \varepsilon_{nj} $ are independently and identically distributed (i.i.d.) according to a Gumbel distribution across agents and alternatives \citep{dmcbook}.

The deterministic utility $ V_{nj} $ is typically specified as a linear-in-parameters function of choice variables plus an alternative-specific constant (ASC):
\begin{equation}
    V_{nj} = \alpha_j + \sum_{k} \beta_{jk} x_{njk},
    \label{eq:deterministic_utility}
\end{equation}
where $ \alpha_j $ is an ASC that accounts for systematic preferences toward mode $ j $ not captured by observed variables, $ x_{njk} $ denotes the value of variable $ k $ (e.g., travel time, cost) associated with alternative $ j $ for agent $ n $, and $ \beta_{jk} $ is the corresponding parameter reflecting the marginal utility or sensitivity to changes in variable $ k $. These parameters are typically estimated from empirical data and calibrated to align with observed mode shares.

The probability that agent $ n $ selects alternative $ j $ from the choice set $ J $ is given by the logit formula:
\begin{equation}
    P_{nj} = \frac{\exp(V_{nj})}{\sum_{m \in J} \exp(V_{nm})}.
    \label{eq:choice_probability}
\end{equation}

This expression defines a probabilistic choice rule where the likelihood of selecting a given mode increases with its relative utility compared to other available options.

In practice, individual choices are realized through stochastic sampling procedures. Two mathematically equivalent methods can be used:

\begin{enumerate}
    \item \textbf{Pseudo-random errors:} A Gumbel-distributed random error term is independently drawn for each alternative. The chosen alternative is the one yielding the highest total utility \citep{horl2021integrating}.
    
    \item \textbf{Probabilistic Sampling Method:} A single draw is made from the multinomial distribution defined by the choice probabilities $ P_{nj} $.
\end{enumerate}

Both approaches preserve the stochastic nature of individual decisions, preventing deterministic convergence and allowing for behavioral variation across agents and iterations. However, differences in simulation dynamics and convergence behavior may arise depending on how stochasticity is managed within the model. In the pseudo-random errors, for instance, if the random error terms are drawn once per agent at the beginning of the simulation and remain fixed thereafter, identical utility values will consistently lead to the same mode choice across iterations. This mechanism can be interpreted as a form of behavioral inertia, capturing the tendency of individuals to persist in their travel mode preferences over time, even when alternatives offer similar utilities. 

\subsection{Stochastic Gradient Methods}

The concept of updating model parameters using a subset of noisy observations is fundamental in ML, particularly when dealing with large datasets. Stochastic Gradient (SG) methods are widely used to adjust model parameters iteratively by computing the gradient of the loss function on small, randomly selected subsets of data called mini-batches \citep{bottou2018optimization,gradientDescent}. This allows the model to gradually improve without processing the entire dataset at once.

The optimization advances iteratively toward the minimum following this general equation:
\begin{equation}
    w_{t+1} = w_t - \gamma_t \nabla_w \mathcal{L}(w_t; \mathcal{B}_t),
\end{equation}
where $w_t$ represents the model parameters (weights in ML) at iteration $t$, $\gamma_t$ is the learning rate (which can be constant or variable), and $\nabla_w \mathcal{L}(w_t; \mathcal{B}_t)$ is the gradient of the loss function computed on mini-batch $\mathcal{B}_t$. At each step, the parameters are updated in the direction that reduces the error, but the size of the update is controlled by the learning rate, ensuring changes are small and stable rather than abrupt.

To address the inherent noise in stochastic gradients, several enhancements have been developed. Momentum methods \citep{nakerst2020gradient, bottou2018optimization} incorporate information from previous updates to smooth the optimization trajectory:
\begin{equation}
    v_{t+1} = \psi v_t + \gamma_t \nabla_w \mathcal{L}(w_t; \mathcal{B}_t), \quad w_{t+1} = w_t - v_{t+1},
\end{equation}
where $\psi$ controls the influence of past gradients. This helps the algorithm maintain consistent progress through flat regions and reduces oscillations in narrow valleys of the loss landscape.

Modern variants like RMSprop and Adam \citep{adam_optimizer} further adapt the learning rate for each parameter individually, leading to more robust convergence across diverse optimization landscapes. These methods have become essential tools in training deep neural networks, where the combination of large datasets and complex parameter spaces makes conventional gradient-based nonlinear optimization techniques impractical.

\section{Methodology}
\label{sec:methodology}

This section presents the proposed framework for parameter calibration. To illustrate the methodology, we focus on calibrating the parameters of the utility functions used in discrete mode choice in MATSim ($\alpha_j$ and $\beta_{jk}$ in \autoref{eq:deterministic_utility}). However, this approach is not limited to these parameters, other simulation parameters, such as link capacities and free-flow speeds can be calibrated using the same methodology with minor modifications.

\subsection{Proposed Calibration Framework}

The methodology builds upon a key feature of MATSim's discrete mode choice extension: at each iteration, only a small fraction of agents (typically 5\%) undergo replanning and potentially change their mode choices. This selective replanning leads to minimal changes in network conditions, such as link travel times or congestion levels, between consecutive iterations, particularly after an initial warm-up phase (first few MATSim iterations). As a result, the decision context faced by replanning agents remains approximately constant between two consecutive MATSim iterations.

We formalize this observation through the \textit{quasi-static assumption}: the choice variables (e.g., travel time, cost, distance) for a given agent and trip and mode vary negligibly between two successive MATSim iterations. Under this assumption, the choice variables experienced by agents during replanning can be recorded and reused for off-line parameter optimization.

Specifically, during the execution of MATSim, we systematically log all the choice variables for all agents who perform mode choice replanning in MATSim iteration $i$. Let $\mathbf{x}_{nmt}^{(i)}$ denote the vector of choice variables for agent $n$, trip $t$, and mode $m$ at MATSim iteration $i$. These variables are stored along with the agent's chosen mode $y_{nt}^{(i)} \in \mathcal{M}$, where $\mathcal{M}$ is the set of available modes.

This data collection produces a dataset $\mathcal{D}^{(i)} = \{(\mathbf{x}_{nmt}^{(i)}, y_{nt}^{(i)})\}$ of choice situations and outcomes, which serves as input for the subsequent parameter optimization stage. The dataset enables direct computation of the deterministic mode utility for each agent and alternative using \autoref{eq:deterministic_utility}, while choice probabilities are estimated via \autoref{eq:choice_probability}.

Let $\theta = \{\alpha_j, \beta_{jk}\}$ denote the candidate parameter vector. Given the stored dataset $\mathcal{D}^{(i)}$, the expected iteration outcomes for the corresponding set of agents can be computed for any such parameter vector by applying \autoref{eq:deterministic_utility} and sampling mode choices for each agent and each trip. This framework enables the deployment of optimization algorithms to identify optimal parameters $\theta^*$ that minimize an objective function quantifying the discrepancy between observed real-world data and simulation outputs.

The choice of objective function depends on data availability. When aggregate statistics are available, the optimization targets distributional measures such as mode shares and their spatial or temporal distributions. Alternatively, when individual choice observations are available (e.g., survey agents injected into the simulation), maximum likelihood estimation can be employed to calibrate model parameters based on observed individual decisions.

\begin{figure}[h]
    \centering
    \includegraphics[width=0.6\linewidth]{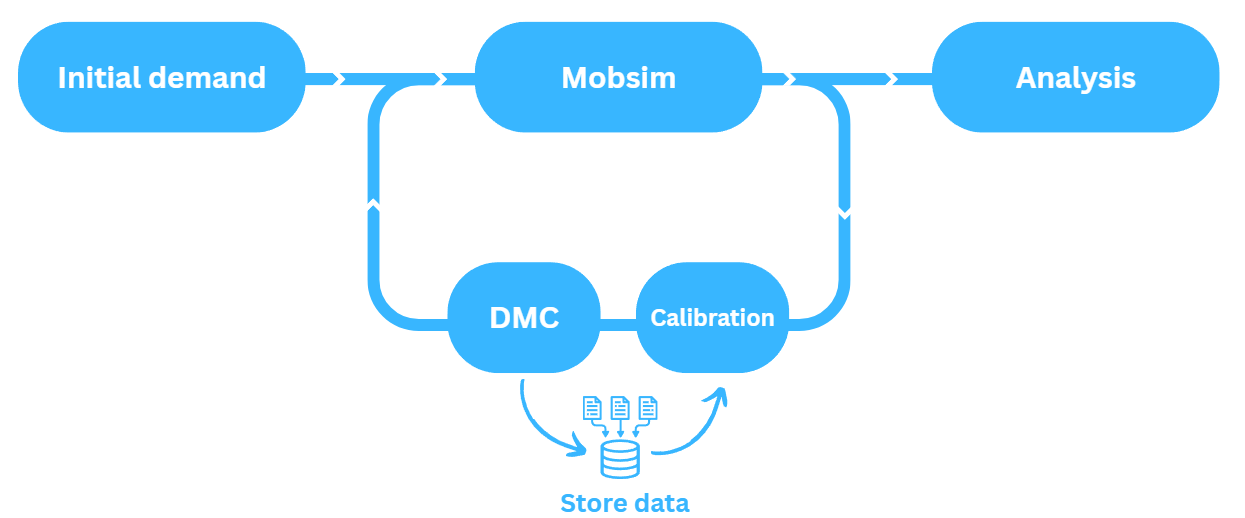}
    \caption{Discrete Mode Choice Calibration Loop}
    \label{fig:dmc_calibration}
\end{figure}

The optimization objective is thus to minimize the discrepancy between simulated and target outcomes, iteratively refining parameter estimates as network conditions evolve through MATSim iterations, until convergence is achieved, as illustrated in \autoref{fig:dmc_calibration}.

\subsection{Optimization}

With choice variables pre-computed and stored (in $\mathcal{D}^{(i)}$), utility evaluation becomes computationally efficient, facilitating the integration of various derivative-free optimization algorithms. Our implementation employs a modular Python framework that supports multiple optimizers, including CMA-ES, SPSA, PSO, Tree-structured Parzen Estimator (TPE), and Genetic Algorithm (GA).\\

\subsubsection{Problem formulation}

The optimization problem at each MATSim iteration $i$ can be mathematically formulated as:
\begin{equation}
    \theta^* = \arg\min_{\theta} \mathcal{L}(\theta) = \arg\min_{\theta} f\left(\mathbf{y}^{obs}, \mathbf{y}^{(i)}(\theta)\right),
\end{equation}
where $\mathcal{L}(\theta)$ represents the objective function and $f(\mathbf{y}^{obs}, \mathbf{y}^{(i)}(\theta))$ is the loss function that measures the discrepancy between observed and simulated outcomes. The vector $\mathbf{y}^{obs}$ denotes the observed data (e.g., mode shares, travel patterns), while $\mathbf{y}^{(i)}(\theta)$ represents the simulated outcomes for the set of agents that replanned their activities in MATSim iteration $i$. These outcomes would normally be generated by MATSim, but are here estimated offline using the dataset $\mathcal{D}^{(i)}$.

For instance, in agents' mode decisions, for each agent $n$ and trip $t$, the deterministic utility $V_{nmt}^{(i)}(\theta)$ is calculated using \autoref{eq:deterministic_utility} for each available mode $m$. The agent's mode choice is then determined by:
\begin{equation}
    y_{nt}^{(i)} = \arg\max_{m \in \mathcal{M}_{nt}^{(i)}} \left[ V_{nmt}^{(i)}(\theta) + \varepsilon_{nmt} \right]
\end{equation}

where $\varepsilon_{nmt} \sim \text{Gumbel}(0,1)$ are independent random error terms following a Gumbel distribution, and $\mathcal{M}_{nt}^{(i)}$ represents the feasible choice set at MATSim iteration $i$.

The choice set $\mathcal{M}_{nt}^{(i)}$ is iteration-dependent to ensure compliance with the availability constraints implimented in MATSim, such as vehicle availability requirements. For instance, if an agent uses a private vehicle for the morning trip, they must use the same vehicle for the return trip and cannot leave it at the destination. Conversely, if an agent did not use their private vehicle for the morning journey, it remains unavailable for the return trip, thereby constraining the feasible mode choices throughout the day.\\

\subsubsection{Optimization Process}
The optimization process operates as follows: optimization algorithms propose parameter sets $\theta$, from which deterministic utilities are computed using \autoref{eq:deterministic_utility}. During the initial iteration of the optimization, random error terms $\varepsilon_{nmt}$ are sampled from a Gumbel distribution. These error terms are added to the deterministic utilities, and the mode with the highest total utility is selected for each agent (Method 1 in \autoref{sec:dmc}). 

This sampling approach is adopted to ensure the loss function remains deterministic during one optimization, thereby simplifying the optimization. Stochastic sampling is unnecessary since optimization is applied iteratively to random subsets of agents (through MATSim iterations). Consequently, the optimal parameters identified in MATSim iteration $i$ serve primarily to determine the direction of parameter updates rather than to provide final optimal parameter values.

The resulting choice outcomes $\mathbf{y}^{(i)}(\theta)$ are used to evaluate the loss function $\mathcal{L}(\theta)$, which is then fed back to the optimizer. The optimizer proposes new parameter sets based on this feedback, and this process iterates until convergence criteria are met:
$$|\mathcal{L}(\theta^{(k+1)}) - \mathcal{L}(\theta^{(k)})| < \epsilon,$$

or a maximum number of iterations is reached, ultimately identifying the optimal parameter set $\theta^*$.\\

\subsubsection{Calibration objectives}
The optimization process minimizes a loss function ($f$) that varies according to the problem specification and available data. For utility parameters calibration, several loss functions are implemented in the framework, including statistical measures such as Mean Squared Error (MSE) and Mean Absolute Error (MAE), as well as probabilistic divergence measures including Jensen-Shannon (JS) divergence and Kullback-Leibler (KL) divergence.

When detailed survey data are available, an alternative approach involves including surveyed individuals as virtual agents in the simulation. Since the actual choices of these agents are known, discrete choice models can be re-estimated within each MATSim iteration using specialized software such as Biogeme \citep{biogeme}.\\

\subsubsection{Special Case: Mode-Specific Constants Calibration}
\label{sec:specialCase}
For the special case where only ASCs require calibration to align the simulated mode shares with the observed mode shares, we incorporate the closed-form update rules proposed by \cite{berlinFormula}, with minor adjustments. Since utilities are relative, one of the ASCs must be fixed during calibration to ensure solution uniqueness. The mode with the fixed constant is designated as reference mode "$o$", while all other modes are indexed as "$j$". This calibration approach provides a computationally efficient update mechanism that does not require utility re-estimation or an optimization. Given parameters from MATSim iteration $i-1$, the optimal parameters for iteration $i$ are calculated as:

\begin{equation} \label{equa:nagelEquation}
    \alpha_{j}^{*(i)} = \alpha_{j}^{(i-1)} + (\ln(z_j) - \ln(m_j^{(i)})) - (\ln(z_o) - \ln(m_o^{(i)})).
\end{equation}

Where $\alpha_{j}^{*(i)}$ represents the optimal constant for mode $j$ at MATSim iteration $i$, $\alpha_{j}^{(i-1)}$ denotes the constant for mode $j$ from MATSim iteration $i-1$, $m_j^{(i)}$ is the simulated mode share of mode $j$ among agents who replanned at MATSim iteration $i$, $z_j$ is the observed mode share of mode $j$, and $z_o$ and $m_o^{(i)}$ refer to the reference mode with the fixed constant.

The difference between Equation~\eqref{equa:nagelEquation} and the original formula in \cite{berlinFormula} is that $m_j^{(i)}$ represents the simulated mode share of mode $j$ among all agents in the original formulation, whereas here it is restricted to only those agents who replanned their plans at MATSim iteration $i$.

\subsection{Parameters update}

Abrupt variations in the parameter vector $\theta$ may occur between MATSim iterations, since optimization is performed on subsets of agents and network conditions can vary considerably at the local level. To reduce these variations, we introduce a momentum-based mechanism that smooths parameter updates and improves convergence stability. Several techniques can achieve this smoothing effect. For example, the difference between the optimal parameters proposed at MATSim iteration $i$, $\theta^{*(i)}$, and the parameters applied at the previous MATSim iteration, $\theta^{(i-1)}$, i.e. $(\theta^{*(i)} - \theta^{(i-1)})$, can be interpreted as a \textit{pseudo-gradient}, enabling the application of advanced optimization algorithms from ML field, such as RMSProp or Adam.

As a simple and efficient alternative, we adopt an exponential moving average (EMA) update rule. The parameter vector $\theta^{(i)}$ applied at MATSim iteration $i$ is given by:

\begin{equation} \label{equa:ema}
\theta^{(i)} = \gamma \theta^{(i-1)} + (1 - \gamma) \theta^{*(i)},
\end{equation}
where $\theta^{*(i)}$ denotes the raw optimal parameters proposed by the optimizer, $\theta^{(i-1)}$ are the smoothed parameters from the previous MATSim iteration, and $\gamma \in [0,1)$ is the decay factor. This approach suppresses parameter oscillations and stabilizes the learning process.

\section{Experiments} 
\label{sec:experiments}

This section presents a real-world application of the proposed calibration framework, illustrating its ability to align simulated mode choice behavior with observed travel patterns in one simulation run. The framework is applied to a large-scale agent-based model of Switzerland \citep{switzerlandScenario}, implemented within the eqasim pipeline. The Switzerland scenario is particularly suitable for this purpose, as it covers a diverse mix of urban, suburban, and rural mobility patterns, providing a comprehensive testbed for evaluating calibration performance.

The Switzerland scenario is publicly available and has been extensively validated in transportation research applications \citep{horl2021simulation}. Although the scenario framework is open-source, certain input datasets remain subject to institutional usage restrictions and are not publicly accessible. The methodological framework presented herein can alternatively be applied to the Île-de-France scenario \citep{idfEqasim}, which provides unrestricted access.

The Switzerland model covers the entire national territory across all cantons and employs the discrete mode choice extension of MATSim. To enhance computational efficiency, large-scale agent-based simulations typically employ a representative sample of the population \citep{llorca2019effects}. This study implements a 5\% sample of the total population, which provides sufficient resolution for capturing system-wide behavioral patterns while maintaining computational efficiency.

The model explicitly considers five transportation modes: walking, cycling, car (as driver), car passenger, and public transport (PT). Car passengers, however, do not make mode choices and are assumed to remain passengers throughout the simulation. Deterministic utilities for the other four modes are defined in equations~(\ref{eq:utilityCar}--\ref{eq:utilityCp}) \citep{dmc1}.

\begin{align}
    % --- Car Utility ---
    \label{eq:utilityCar}
    u_{\text{car}}(\chi) &= \alpha_{\text{car}} + \beta_{\text{travelTime,car}} \cdot \chi_{\text{travelTime,car}} \nonumber \\
    &\quad + \beta_{\text{travelTime,walk}} \cdot \theta_{\text{accessEgressWalkTime}} \nonumber \\
    &\quad + \beta_{\text{cost}} \cdot \left( \frac{\chi_{\text{crowflyDistance}}}{\theta_{\text{averageDistance}}} \right)^\lambda \cdot \chi_{\text{cost,car}} \\    
    %
    % --- Public Transport Utility ---
    u_{\text{pt}}(\chi) &= \alpha_{\text{pt}}  + \beta_{\text{accessEgressTime}} \cdot \chi_{\text{accessEgressTime}} \nonumber \\
    &\quad + \beta_{\text{inVehicleTime}} \cdot \chi_{\text{inVehicleTime}} \nonumber \\
    &\quad + \beta_{\text{waitingTime}} \cdot \chi_{\text{waitingTime}} \nonumber \\
    &\quad + \beta_{\text{numberOfTransfers}} \cdot \chi_{\text{numberOfTransfers}} \nonumber \\
    &\quad + \beta_{\text{cost}} \cdot \left( \frac{\chi_{\text{crowflyDistance}}}{\theta_{\text{averageDistance}}} \right)^\lambda \cdot \chi_{\text{cost,pt}} \\
    %
    % --- Bike Utility ---
    u_{\text{bike}}(\chi) &= \alpha_{\text{bike}} + \beta_{\text{travelTime,bike}} \cdot \chi_{\text{travelTime,bike}} +\beta_{\text{age,bike}} \cdot \max(0, a_{\text{age}} - 18) \\
    %
    % --- Walk Utility ---
    u_{\text{walk}}(\chi) &= \alpha_{\text{walk}} + \beta_{\text{travelTime,walk}} \cdot \chi_{\text{travelTime,walk}} 
    \label{eq:utilityCp}
\end{align}

The following sections present three experiments. In each experiment, we calibrate a different set of parameters. The first experiment uses the overall approach proposed in this paper, while the other two calibrate only the mode-specific constants, using the special-case formula presented in \autoref{sec:specialCase}.

\subsection{Calibration of Mode-Specific Constants and Travel Time Marginal Utilities}

This section presents a first application of our proposed methodology, focusing on the simultaneous calibration of seven utility parameters: mode-specific constants $(\alpha_{car}, \alpha_{bike}, \alpha_{walk})$ and travel time marginal utilities $(\beta_{travelTime,car}, \beta_{travelTime,walk}, \beta_{travelTime,bike},\beta_{inVehicleTime,pt})$ from equations~(\ref{eq:utilityCar}--\ref{eq:utilityCp}), while the remaining parameters are kept at their default values used in the model.

Given the high dimensionality and complexity of this optimization problem, we employ the CMA-ES optimizer, with a population size of 40. The decay rate ($\gamma$ in \autoref{equa:ema}) is initially set to 0.5 and increased every 30 MATSim iterations by halving the gap to 1.

The selection of CMA-ES was based on comparative evaluation of several optimization algorithms. Alternative methods either failed to converge or exhibited lower computational efficiency for the given problem formulation. To improve computational performance, a warm-start strategy was implemented: during the first MATSim iteration, CMA-ES completes the full optimization. In subsequent MATSim iterations, it initializes using previously evaluated solutions and their objective values. This approach reduces the number of optimizer iterations needed for convergence by leveraging historical search information.

The objective function employs the logarithm of the MSE as the loss measure to minimize discrepancies between observed and simulated mode shares at two levels: global mode shares, and the distribution of mode shares as a function of trip euclidean distance. This dual-level approach ensures that the calibrated parameters capture both aggregate modal preferences and distance-dependent modal behavior. The observed mode shares are obtained from the Swiss Microcensus on Mobility and Transport (Mikrozensus Mobilität und Verkehr \cite{microcensus}).

\begin{figure}[h]
    \centering
    \includegraphics[width=0.55\linewidth]{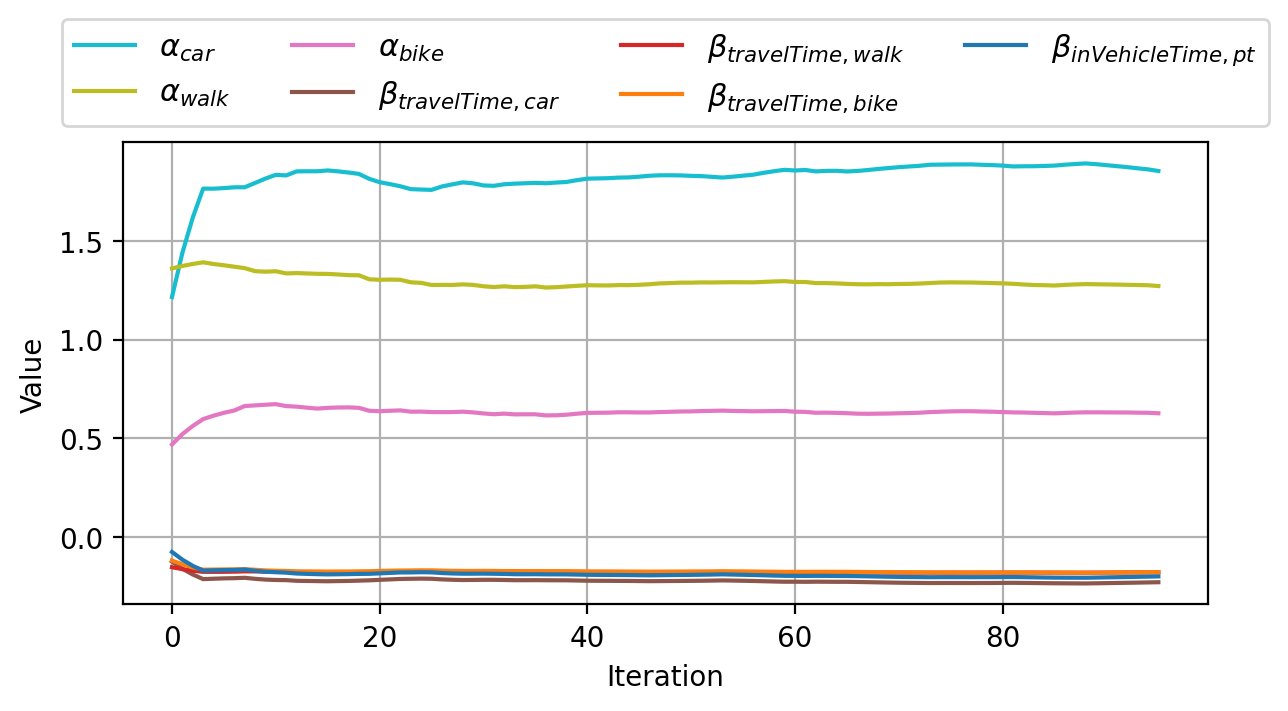}
    \caption{Evolution of calibrated parameters across MATSim iterations}
    \label{fig:paramsEvol}
\end{figure}

\begin{figure}[h]
    \centering
    \begin{minipage}[h]{0.48\linewidth}
        \centering
        \vspace{-1pt}
        \includegraphics[width=0.95\linewidth]{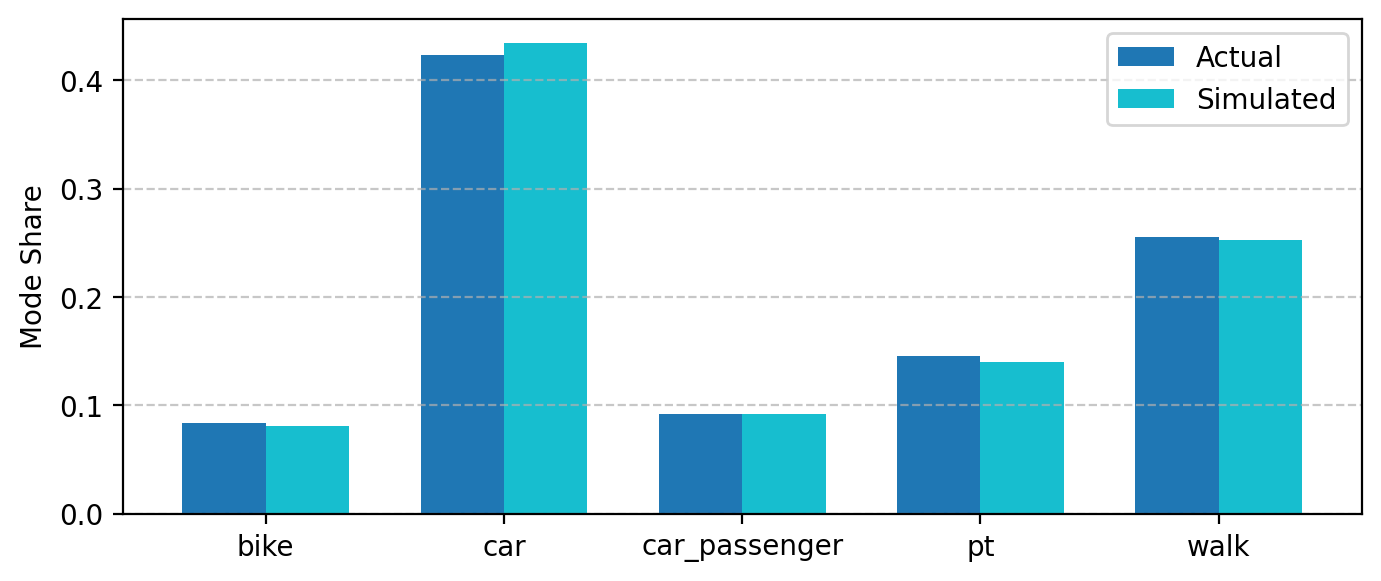}   
        \vspace{1pt}
        \captionof{figure}{Global mode shares for the simulated modes}
        \label{fig:paramsShares}
    \end{minipage}%
    \hfill
    \begin{minipage}[h]{0.48\linewidth}
        \centering
        \includegraphics[width=\linewidth]{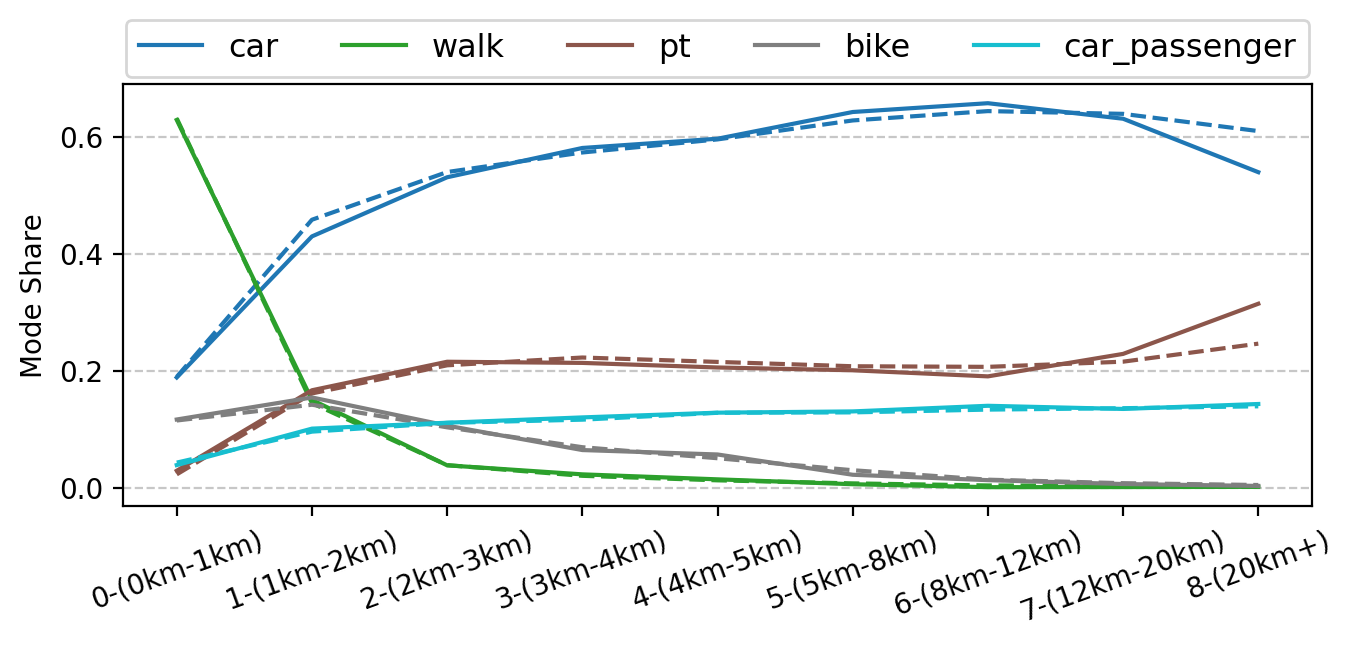}
        \captionof{figure}{Distribution of mode shares with respect to Euclidean distance for the simulated modes}
        \label{fig:paramsDist}
    \end{minipage}
\end{figure}

The calibration results are presented in three figures: \autoref{fig:paramsEvol} shows the evolution of calibrated parameters across MATSim iterations, \autoref{fig:paramsShares} compares simulated global mode shares with observed mode shares, and \autoref{fig:paramsDist} presents the distribution of mode shares as a function of trip euclidean distance for all five transport modes. 

Parameter values converge rapidly during the initial MATSim iterations and then stabilize. The final calibrated values are computed as the mean over the last few iterations.

Mode shares comparison reveals good agreement between simulated and observed mode shares. Small discrepancies, mainly in the distribution, can be attributed to the partial calibration approach, where only a subset of utility parameters are adjusted. Achieving perfect alignment would require either expanding the calibration to include additional parameters or focusing on those with the greatest impact.

\subsection{Mode Specific Constants Calibration}
In the second calibration test, we adjust only the ASCs of the utilities. Since utilities are defined relative to each other, one ASC must be fixed. We set $\alpha_{pt}$ to zero, so that only $\alpha_{car}$, $\alpha_{walk}$ and $\alpha_{bike}$ are calibrated. Three decay rates ($\gamma$) are tested: 0.3, 0.5, and 0.7. For each case, the decay rate is gradually increased during MATSim iterations to approach a value of 1, thereby reducing parameter fluctuations.

Because only the ASCs are calibrated, there is no need to use a complex optimization algorithm. Instead, \autoref{equa:nagelEquation} is applied after each iteration to obtain the optimal ASC values. Once the optimal values are computed, \autoref{equa:ema} is used to smooth the variations.

The objective is to align the global simulated mode shares with the observed global mode shares. The observed mode shares are obtained from the Swiss Microcensus on Mobility and Transport.

\begin{figure}[h]
    \centering
    \includegraphics[width=0.55\linewidth]{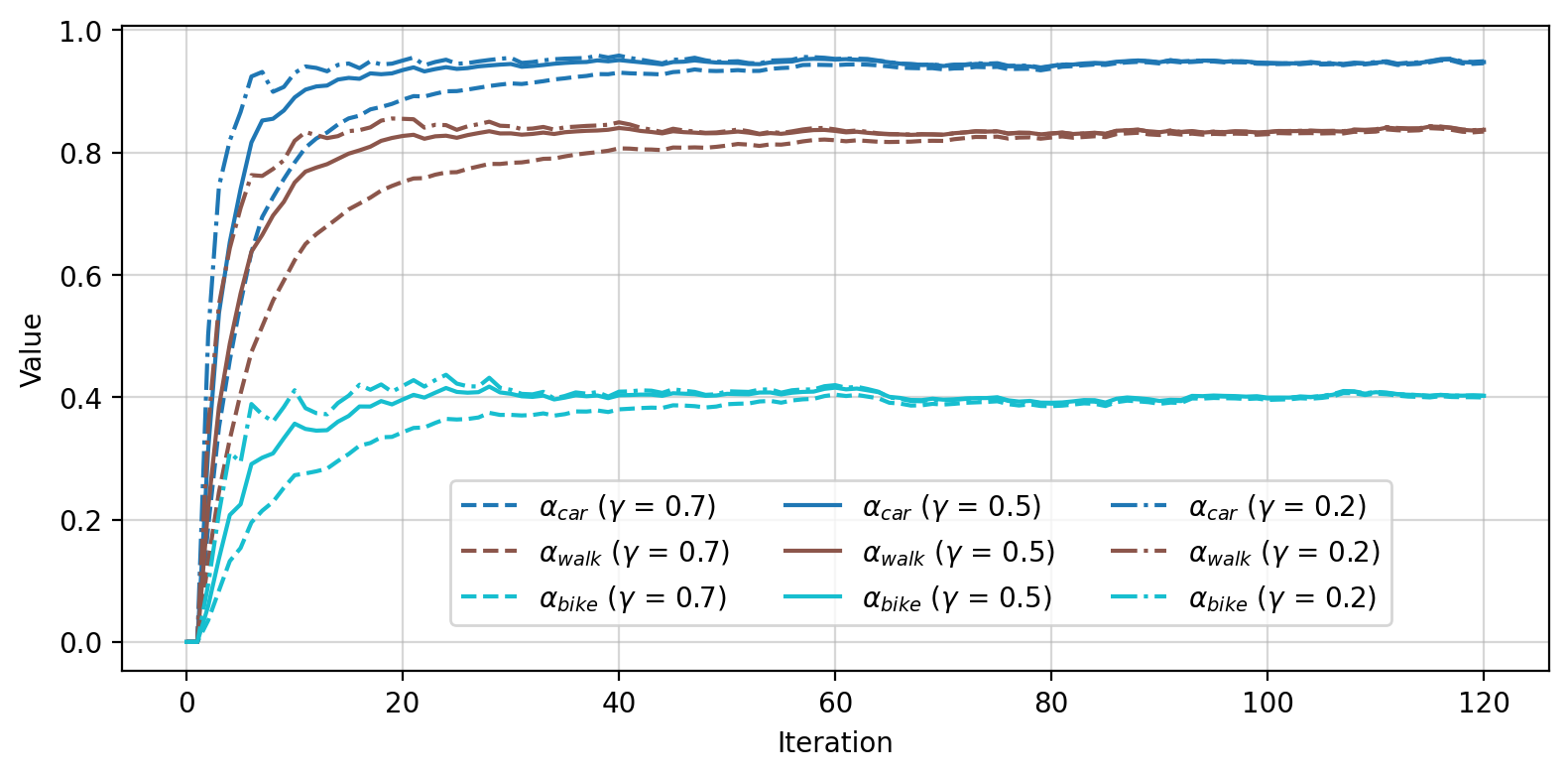}
    \caption{Evolution of the mode-specific constants $\alpha_j$ across MATSim iterations}
    \label{fig:alphaCalibEvol}
\end{figure}

\begin{figure}[h]
    \centering
    \includegraphics[width=0.55\linewidth]{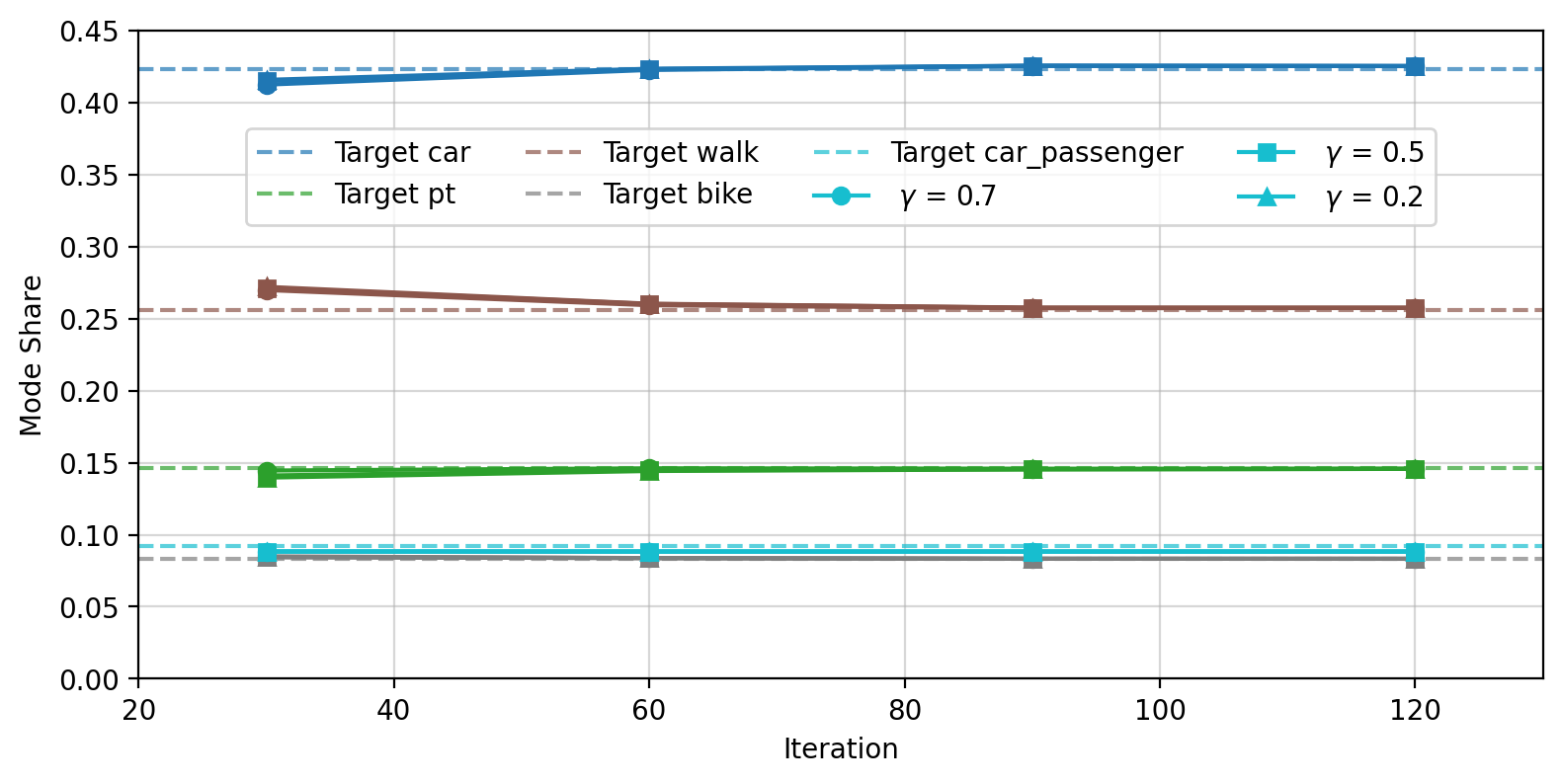}
    \caption{Mode shares across different MATSim iterations}
    \label{fig:AlphaCalibShares}
\end{figure}

The simulation starts with all ASCs values set to zero and runs for 120 iterations. \autoref{fig:alphaCalibEvol} shows the evolution of the ASCs across MATSim iterations, while \autoref{fig:AlphaCalibShares} displays the mode shares at MATSim iterations 30, 60, 90, and 120 for the three experiments and for all simulated modes.

All experiments converge to the same optimal parameters. Mode shares are already matched within 60 MATSim iterations for all three cases. The only difference lies in the convergence rate: the experiment with the lowest $\gamma$ converges faster. After 60 iterations, the parameters remain stable across all experiments, as the optimal values have been reached and $\gamma$ is already high due to its systematic increase, which reduces fluctuations between MATSim iterations.

\subsection{Canton-Specific Mode Specific Constant Calibration} \label{sec:cantonsCalibration}

Switzerland consists of 26 cantons that exhibit considerable heterogeneity in language, taxation, geography, and transportation infrastructure. These regional differences influence mode choice decisions through factors not explicitly modeled in the simulation and are accounted for through mode-specific constants. While a single national mode-specific constant can accurately reproduce aggregate mode shares, it fails to capture local variations, resulting in significant discrepancies at the cantonal level.

To address this limitation, we implement a canton-specific calibration approach in which mode-specific constants vary according to the agent's canton of residence. The mode-specific constants $\alpha_j$ in equations~(\ref{eq:utilityCar}--\ref{eq:utilityCp}) are replaced with $\alpha_{jc}$, where $c$ denotes the agent's canton of residence.

For calibrating $\alpha_{jc}$, we employ \autoref{equa:nagelEquation}, adapted for this application. This equation is applied within each canton by adjusting parameters $m$ and $z$ to reflect mode shares exclusively for residents of that canton. The method determines canton-specific constants in each MATSim iteration based on the mode choice decisions of agents residing in the respective canton. This approach assumes independence between cantons, which constitutes a simplification given that inter-cantonal mobility patterns create spatial dependencies. However, these interactions exhibit limited variation between consecutive MATSim iterations, allowing the $\alpha_{jc}$ parameters across all cantons to converge simultaneously to their optimal values. In scenarios where inter-cantonal mobility effects become significant, the decay rate can be increased to ensure stable convergence of the calibration parameters.

Given the use of a 5\% population sample and the fact that only 5\% of agents replan their trips in each MATSim iteration, the number of replanning agents per canton per MATSim iteration may be low, potentially resulting in high parameter fluctuations between iterations. To address this issue, we implement a batch size threshold mechanism.

The calibration procedure operates as follows: in each MATSim iteration and canton, parameters are updated only if the number of agents that replanned their trips exceeds a predefined threshold. If the threshold is not met, parameter updates are deferred. For cantons with smaller populations, replanning agents from successive iterations are accumulated until the threshold is reached, at which point the parameters for that canton are updated. In this study, the threshold is set to 800 agents. 

\begin{figure}[h]
    \centering
    \includegraphics[width=0.6\linewidth]{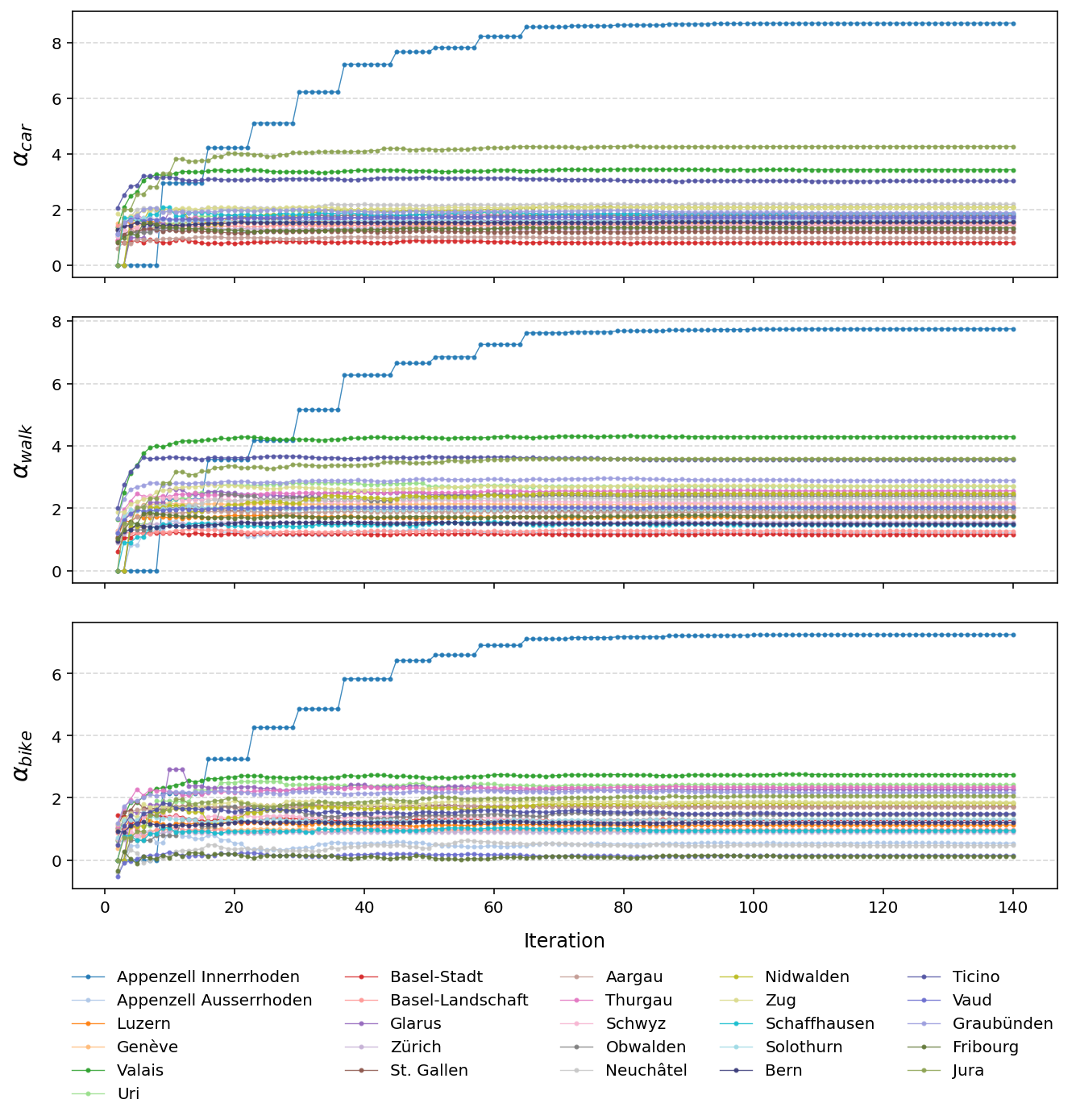}
    \caption{Evolution of calibrated parameters across MATSim iterations}
    \label{fig:cantonsParamsEvolution}
\end{figure}

All initial $\alpha_{jc}$ values are set to zero, and the initial decay rate ($\gamma$) is set to 0.2 to accelerate convergence in the first iterations and is gradually increased over successive MATSim iterations, approaching 1.

The evolution of the calibrated parameters across MATSim iterations is depicted in \autoref{fig:cantonsParamsEvolution}. The figure shows that not all parameters are updated in each MATSim iteration, some parameters are updated only once every multiple iterations. These parameters belong to cantons with low population. This effect can be controlled by adjusting the batch size threshold.

Most parameters stabilize before iteration 60, with the exception of the canton of Appenzell Innerrhoden. Because this canton has a very small population, parameter updates occur only after several iterations. In addition, the target public transport mode share is very low, which requires more adjustments to reach convergence. The calibrated parameters for this canton may be overfitted. However, without additional data on actual mode shares, it is not possible to validate or refine the calibration further.

\begin{figure}[h]
    \centering
    \includegraphics[width=0.6\linewidth]{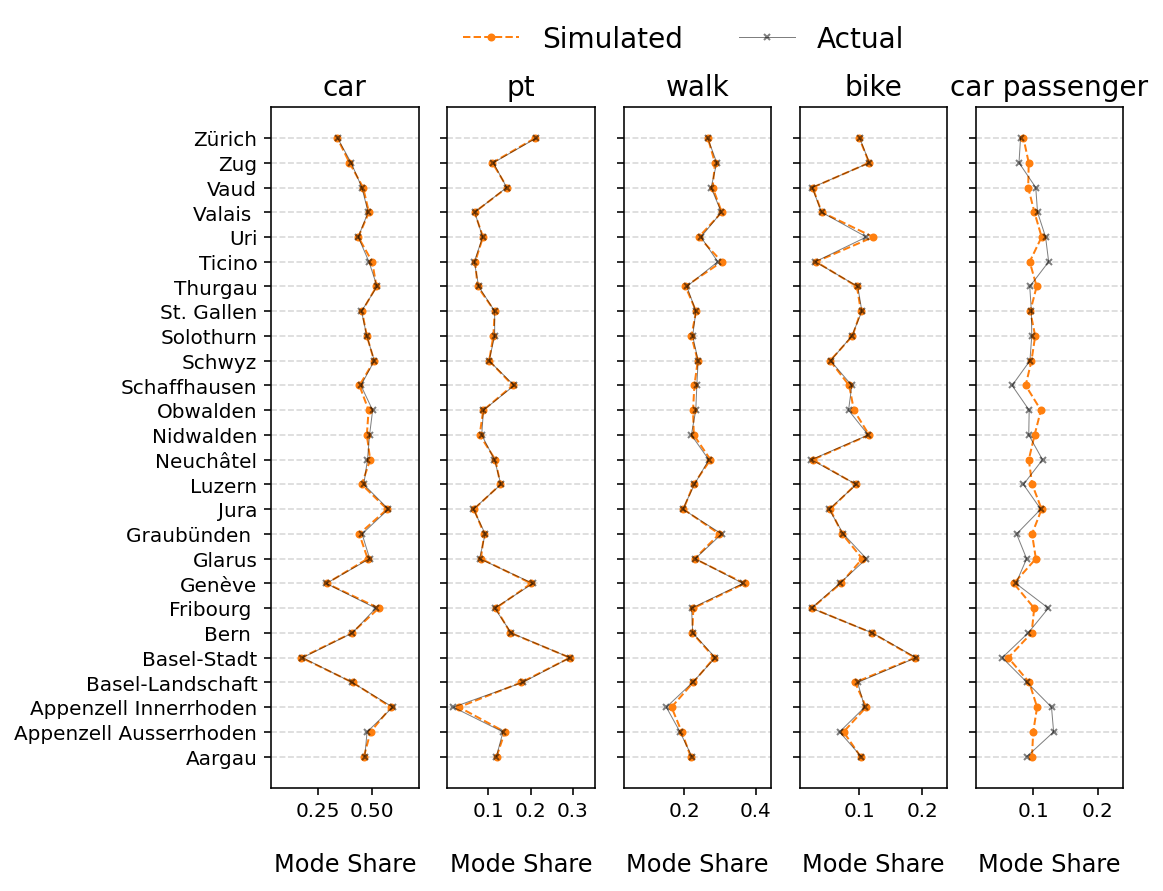}
    \caption{Mode shares within each canton}
    \label{fig:cantonsModeShare}
\end{figure}

The mode share results for the five modes by canton are depicted in \autoref{fig:cantonsModeShare}. The calibrated parameters allow reproduction of similar mode shares within each canton. The highest discrepancies are observed for car passengers, who are not permitted to perform mode choice and remain fixed to their initial choices assigned during population creation in eqasim. These differences can be partly explained by the use of a 5\% population sample. This is supported by the observation that higher population cantons (e.g., Zurich, Bern, Aargau) show smaller discrepancies. The small deviations in mode shares for the other modes (car, public transport, walking, cycling) compensate for the discrepancies in the car passenger category.

\section{Discussion}
\label{sec:discussion}
The proposed online calibration approach successfully aligned simulated mode shares with observed values within a single simulation run. Calibration converged in approximately 60 iterations, which is comparable to the number of iterations required without calibration.

\subsection{Computational Performance}

When calibrating only mode-specific constants, the computational overhead is negligible, requiring the same runtime as uncalibrated simulations. For more comprehensive parameter calibration using black-box optimization, runtime increases depend on several factors: the chosen optimizer, optimization parameters (evaluation count, tolerance), and the number of calibrated parameters. Our experiments showed optimization times of 2-3 minutes per MATSim iteration, which represents an acceptable computational burden.

During early MATSim iterations, optimization convergence criteria can be relaxed to enable faster parameter updates. Since network conditions continue evolving, precise optimal values are less important than identifying the correct parameter adjustment direction for subsequent MATSim iterations.

\subsection{Methodological Limitations and Challenges}

A significant limitation in existing optimization-based calibration literature (presented in \autoref{sec:relatedWork}) is the absence of repeated trials of the optimization to assess parameter stability. This concern is particularly relevant given the strong correlations among both objectives (travel time, distance, cost, transfers) and parameters, which often create compensatory effects and non-unique solutions if the objective function is not well defined. These result in complex, non-convex optimization landscapes with multiple local minima, making parameter robustness difficult to evaluate without multiple optimization runs from different starting points.

Our online calibration approach presents unique challenges in this context. Unlike traditional methods that rely on repeated independent simulation runs to identify optimal solutions, our method continuously updates parameters during one simulation. In this setting, the issue of multiple local minima becomes particularly critical: optimizers may converge to different solutions across MATSim iterations, preventing parameter stabilization. When combined with momentum-based parameter smoothing, this effect can trap parameters in regions that represent mixtures of different local minima rather than true optima. Such regions may not be optimal.

From a behavioral standpoint, there is no guarantee that the calibrated parameters are behaviorally realistic. The analyst may therefore need to impose bounds on parameter values to ensure behavioral plausibility. This paper focuses on the methodological aspects, and addressing behavioral validity lies beyond its scope. However, this issue remains crucial, as unrealistic parameter values may compromise the interpretability and applicability of the results in practice.

\subsection{Recommendations for Implementation}

To address optimization challenges, we recommend three strategies. First, ensure objective function convexity through careful formulation, in order to obtain consistent optimization results. Second, when convexity cannot be achieved, adjust optimizer parameters (e.g., increasing the CMA-ES population size) to improve the likelihood of convergence to the same global optimum across iterations. Third, when convergence to a single global optimum is not required, adopt a progressive optimization approach: perform a full optimization in the initial MATSim iteration, and in subsequent iterations resume the optimization using information from previous iterations to initialize (warm-start) the optimizer.

The third option can be implemented in different ways. One effective approach is to aggregate data from the current MATSim iteration together with data from the previous $n$ iterations, thereby increasing the number of data points and reusing part of the data from earlier optimizations. The optimizations are then linked across iterations through this shared data, the use of a consistent random seed, and the incorporation of optimization history from past MATSim iterations to initialize the optimizer in the current iteration.

For highly stochastic agent-based models, including the standard MATSim implementation, the calibration frequency can be reduced by performing optimization every $n$ iterations instead of at every iteration. In this batch approach, the last $n$ iterations are considered jointly, while only the most recent replanning decision is retained for agents that have replanned multiple times (as in \autoref{sec:cantonsCalibration}). In this context, the methodology presented in the previous sections corresponds to the special case where $n = 1$, analogous to batch size selection in machine learning applications.

\subsection{Perspectives / Future Directions}
We showed that \autoref{equa:nagelEquation} can be employed beyond the framework for which it was originally proposed. First, we applied it within MATSim iterations to calibrate the global mode shares. Second, we extended the same principle to calibrate mode shares within each canton. These applications demonstrate the utility of the formula.

A natural next step is to test its use in calibrating marginal utilities. This could be achieved by grouping the parameters to which these marginal utilities apply into bins, and then applying the formula within each bin, in the same way it was applied within each canton.

Future applications should also explore whether this methodology can be transferred to other agent-based models. Another important direction would be to integrate it into the standard MATSim framework (without DMC). Finally, the approach may be extended to the calibration of network parameters, which would broaden its scope and strengthen its relevance for transport simulation studies.

\section{Conclusion}
\label{sec:conclusion}

This paper presented a new methodology for calibrating agent-based models. Whereas traditional calibration typically requires hundreds or even thousands of simulation runs, our approach calibrates the model in a single run. To achieve this, we proposed an online approach that combines stochastic batch optimization, widely used in ML, with black-box optimization. We demonstrated the method through three experiments, calibrating the utility function parameters of the agent-based transport model MATSim.

The results are promising: calibration was achieved within a single simulation run. This opens the possibility of applying the method to other models and different types of calibration problems.

We also discussed the challenges and possible improvements of this approach. Future work includes identifying suitable objective functions, testing alternative optimization algorithms, and developing more advanced momentum mechanisms. Beyond this, promising perspectives involve extending the method to other agent-based models, experimenting with different sets of parameters, and even calibrating simulation environment parameters, for example, network capacities and speeds in the case of MATSim.

\section*{Data and Code Availability}

All code and data used in this work are openly available. The calibration framework is implemented in \texttt{eqasim-java}, accessible on GitHub\footnote{\url{https://github.com/eqasim-org/eqasim-java/tree/traffic_light}}.  
The optimizers are implemented from scratch in Python and are available online via GitHub\footnote{\url{https://github.com/Dib-AEK/EqasimParametersCalibration}}.  
The Switzerland scenario is also openly available via Github\footnote{\url{https://github.com/eqasim-org/eqasim-switzerland}}.  

Since the Switzerland scenario is based on access-restricted data, the same methodology can be applied to the Île-de-France scenario, which is publicly available through GitHub\footnote{\url{https://github.com/eqasim-org/ile-de-france}}.

%% References

%The citation must be used in following style: \cite{article-minimal} \cite{article-full} \cite{article-crossref} \cite{whole-journal}.

%% Smaller reference entries; keep the section heading at its normal size.
% \renewcommand{\bibfont}{\footnotesize}
% \bibliography{references}
% \bibliographystyle{elsarticle-harv}

\renewcommand{\bibfont}{\footnotesize}
\setlength{\bibsep}{1.5pt}   % or e.g. 2pt for a little breathing room
\bibliography{references}

@article{bonabeau2002,
  title={{Agent-based modeling: Methods and techniques for simulating human systems}},
  author={Bonabeau, Eric},
  journal={Proceedings of the national academy of sciences},
  volume={99},
  number={suppl\_3},
  pages={7280--7287},
  year={2002},
  publisher={National Academy of Sciences}
}

@inproceedings{amodParis,
  title={{Dynamic demand estimation for an AMoD system in Paris}},
  author={H{\"o}rl, Sebastian and Balac, Milos and Axhausen, Kay W},
  booktitle={2019 IEEE Intelligent Vehicles Symposium (IV)},
  pages={260--266},
  year={2019},
  organization={IEEE}
}

@article{reviewAgentBasedTransportModels,
  title={{Agent-based models in urban transportation: review, challenges, and opportunities}},
  author={Bastarianto, Faza Fawzan and Hancock, Thomas O and Choudhury, Charisma Farheen and Manley, Ed},
  journal={European Transport Research Review},
  volume={15},
  number={1},
  pages={19},
  year={2023},
  publisher={Springer}
}

@inproceedings{dib2023two,
  title={A two-stage deep learning based approach for predicting instantaneous vehicle speed profiles on road networks},
  author={Dib, Abdelkader and Sciarretta, Antonio and Balac, Milos},
  booktitle={2023 IEEE 26th International Conference on Intelligent Transportation Systems (ITSC)},
  pages={1636--1642},
  year={2023},
  organization={IEEE}
}

@article{reviewAbmTransportation,
  title={{A review on agent-based technology for traffic and transportation}},
  author={Bazzan, Ana LC and Kl{\"u}gl, Franziska},
  journal={The Knowledge Engineering Review},
  volume={29},
  number={3},
  pages={375--403},
  year={2014},
  publisher={Cambridge University Press}
}

@article{advancingCalibration,
  title={{Advancing calibration for stochastic agent-based models in epidemiology with Stein variational inference and Gaussian process surrogates}},
  author={Robertson, Connor and Safta, Cosmin and Collier, Nicholson and Ozik, Jonathan and Ray, Jaideep},
  journal={arXiv preprint arXiv:2502.19550},
  year={2025}
}

@article{empiricalValidation,
  title={{Empirical validation of agent-based models: Alternatives and prospects}},
  author={Windrum, Paul and Fagiolo, Giorgio and Moneta, Alessio},
  journal={Journal of Artificial Societies and Social Simulation},
  volume={10},
  number={2},
  pages={8},
  year={2007},
  publisher={University of Surrey}
}

@book{matsim,
  title={{The Multi-Agent Transport Simulation MATSim}},
  author={Horni, Andreas and Nagel, Kai and Axhausen, Kay W},
  year={2016},
  publisher={Ubiquity Press}
}

@article{berlin,
  title={{Advancing the MATSim Open Berlin Scenario: Improvements in transport scenario generation and calibration methods}},
  author={Rakow, Christian and Kreuschner, Moritz and Nagel, Kai},
  journal={Transportation Research Procedia},
  volume={86},
  pages={732--739},
  year={2025},
  publisher={Elsevier}
}

@article{berlinFormula,
  title={{Automated calibration of alternative specific con-stants in agent based transport models}},
  author={Rakow, Christian and Nagel, Kai},
  year={2024}
}

@article{CalUncertainty,
  title={{Calibrating agent-based models using uncertainty quantification methods}},
  author={McCulloch, Josie and Ge, Jiaqi and Ward, Jonathan A and Heppenstall, Alison and Polhill, J Gareth and Malleson, Nick},
  journal={Journal of Artificial Societies and Social Simulation},
  volume={25},
  number={2},
  year={2022},
  publisher={SimSoc Consortium}
}

@article{eqasim,
  title={{Introducing the eqasim pipeline: From raw data to agent-based transport simulation}},
  author={H{\"o}rl, Sebastian and Balac, Milos},
  journal={Procedia Computer Science},
  volume={184},
  pages={712--719},
  year={2021},
  publisher={Elsevier}
}

@article{spgm,
  title={{Deep learning approach to predict microscopic pollutant emissions from mesoscopic traffic simulations}},
  author={Dib, Abdelkader and Balac, Milos and Sciarretta, Antonio},
  journal={Transportation Research Part D: Transport and Environment},
  volume={146},
  pages={104791},
  year={2025},
  publisher={Elsevier}
}

@article{liangcalibration,
  title={{Calibration of Agent-based Transport Simulation with SPSA method}},
  author={Liang, Hong},
  year={2018}
}

@article{spsa1,
  title={{Demand calibration of multimodal microscopic traffic simulation using weighted discrete SPSA}},
  author={Oh, Simon and Seshadri, Ravi and Azevedo, Carlos Lima and Ben-Akiva, Moshe E},
  journal={Transportation Research Record},
  volume={2673},
  number={5},
  pages={503--514},
  year={2019},
  publisher={SAGE Publications Sage CA: Los Angeles, CA}
}

@inproceedings{horl2022exploring,
  title={{Exploring accelerated evolutionary parameter search for iterative large-scale transport simulations in a new calibration testbed}},
  author={H{\"o}rl, Sebastian},
  booktitle={10th symposium of the European Association for Research in Transportation (hEART 2022)},
  year={2022}
}

@article{switzerlandScenario,
  title={{Agent-based transport model of Switzerland: A fully automated pipeline based on eqasim and MATSim}},
  author={Dib, Abdelkader and Sallard, Aurore and Balac, Milo{\v{s}}},
  journal={26th Swiss Transportation Research Conference, Ascona, Switzerland},
  year={2026},
  publisher={CSFM, ETH Zurich}
}

@inproceedings{gradientDescent,
  title={{Large-scale machine learning with stochastic gradient descent}},
  author={Bottou, L{\'e}on},
  booktitle={Proceedings of COMPSTAT'2010: 19th International Conference on Computational StatisticsParis France, August 22-27, 2010 Keynote, Invited and Contributed Papers},
  pages={177--186},
  year={2010},
  organization={Springer}
}

@article{llorca2019effects,
  title={{Effects of scaling down the population for agent-based traffic simulations}},
  author={Llorca, Carlos and Moeckel, Rolf},
  journal={Procedia Computer Science},
  volume={151},
  pages={782--787},
  year={2019},
  publisher={Elsevier}
}

@article{biology1,
  title={{Agent-based models in translational systems biology}},
  author={An, Gary and Mi, Qi and Dutta-Moscato, Joyeeta and Vodovotz, Yoram},
  journal={Wiley Interdisciplinary Reviews: Systems Biology and Medicine},
  volume={1},
  number={2},
  pages={159--171},
  year={2009},
  publisher={Wiley Online Library}
}

@article{biology2,
  title={{Agent-based modeling in molecular systems biology}},
  author={Soheilypour, Mohammad and Mofrad, Mohammad RK},
  journal={BioEssays},
  volume={40},
  number={7},
  pages={1800020},
  year={2018},
  publisher={Wiley Online Library}
}

@article{ecology1,
  title={{The role of agent-based models in wildlife ecology and management}},
  author={McLane, Adam J and Semeniuk, Christina and McDermid, Gregory J and Marceau, Danielle J},
  journal={Ecological modelling},
  volume={222},
  number={8},
  pages={1544--1556},
  year={2011},
  publisher={Elsevier}
}

@article{economic1,
  title={{Agent-based economic models and econometrics}},
  author={Chen, Shu-Heng and Chang, Chia-Ling and Du, Ye-Rong},
  journal={The Knowledge Engineering Review},
  volume={27},
  number={2},
  pages={187--219},
  year={2012},
  publisher={Cambridge University Press}
}

@article{economic2,
  title={{Agent-based modeling in economics and finance: Past, present, and future}},
  author={Axtell, Robert L and Farmer, J Doyne},
  journal={Journal of Economic Literature},
  volume={63},
  number={1},
  pages={197--287},
  year={2025},
  publisher={American Economic Association 2014 Broadway, Suite 305, Nashville, TN 37203-2425}
}

@article{sociology1,
  title={{From factors to actors: Computational sociology and agent-based modeling}},
  author={Macy, Michael W and Willer, Robert},
  journal={Annual review of sociology},
  volume={28},
  number={1},
  pages={143--166},
  year={2002},
  publisher={Annual Reviews 4139 El Camino Way, PO Box 10139, Palo Alto, CA 94303-0139, USA}
}

@article{energySystems1,
  title={{Modeling technological change in energy systems--From optimization to agent-based modeling}},
  author={Ma, Tieju and Nakamori, Yoshiteru},
  journal={Energy},
  volume={34},
  number={7},
  pages={873--879},
  year={2009},
  publisher={Elsevier}
}

@article{landUse,
  title={{Agent-based land-use models: a review of applications}},
  author={Matthews, Robin B and Gilbert, Nigel G and Roach, Alan and Polhill, J Gary and Gotts, Nick M},
  journal={Landscape Ecology},
  volume={22},
  number={10},
  pages={1447--1459},
  year={2007},
  publisher={Springer}
}

@article{bayesian1,
  title={{Estimation of agent-based models using sequential Monte Carlo methods}},
  author={Lux, Thomas},
  journal={Journal of Economic Dynamics and Control},
  volume={91},
  pages={391--408},
  year={2018},
  publisher={Elsevier}
}

@article{bayesian2,
  title={{Bayesian estimation of agent-based models}},
  author={Grazzini, Jakob and Richiardi, Matteo G and Tsionas, Mike},
  journal={Journal of Economic Dynamics and Control},
  volume={77},
  pages={26--47},
  year={2017},
  publisher={Elsevier}
}

@inproceedings{genericAlgo1,
  title={{Assisted parameter and behavior calibration in agent-based models with distributed optimization}},
  author={D’Auria, Matteo and Scott, Eric O and Lather, Rajdeep Singh and Hilty, Javier and Luke, Sean},
  booktitle={International Conference on Practical Applications of Agents and Multi-Agent Systems},
  pages={93--105},
  year={2020},
  organization={Springer}
}

@article{genericAlgo2,
  title={{Calibrating agent-based models using a genetic algorithm}},
  author={Canessa, Enrique and Chaigneau, Sergio},
  journal={Studies in Informatics and Control},
  volume={24},
  number={1},
  pages={79--90},
  year={2015}
}

@article{pso1,
  title={{Particle swarm optimization for calibration in spatially explicit agent-based modeling}},
  author={Michels, Alexander and Kang, Jeon-Young and Wang, Shaowen},
  journal={Journal of Artificial Societies and Social Simulation},
  volume={25},
  number={2},
  year={2022}
}

@article{surrogate1,
  title={{Advancing calibration for stochastic agent-based models in epidemiology with Stein variational inference and Gaussian process surrogates}},
  author={Robertson, Connor and Safta, Cosmin and Collier, Nicholson and Ozik, Jonathan and Ray, Jaideep},
  journal={arXiv preprint arXiv:2502.19550},
  year={2025}
}

@article{surrogate2,
  title={{Using machine learning as a surrogate model for agent-based simulations}},
  author={Angione, Claudio and Silverman, Eric and Yaneske, Elisabeth},
  journal={Plos one},
  volume={17},
  number={2},
  pages={e0263150},
  year={2022},
  publisher={Public Library of Science San Francisco, CA USA}
}

@article{surrogate3,
  title={{Validation and calibration of an agent-based model: A surrogate approach}},
  author={Zhang, Yi and Li, Zhe and Zhang, Yongchao},
  journal={Discrete Dynamics in Nature and Society},
  volume={2020},
  number={1},
  pages={6946370},
  year={2020},
  publisher={Wiley Online Library}
}

@article{agarwal2017calibration,
  title={{Calibration of behavioral parameters in an agent-based transport simulation}},
  author={Agarwal, Amit and Fl{\"o}tter{\"o}d, Gunnar and Nagel, Kai},
  year={2017}
}

@article{flotterod2017search,
  title={{A search acceleration method for optimization problems with transport simulation constraints}},
  author={Fl{\"o}tter{\"o}d, Gunnar},
  journal={Transportation Research Part B: Methodological},
  volume={98},
  pages={239--260},
  year={2017},
  publisher={Elsevier}
}

@article{idfEqasim,
  title={{Synthetic population and travel demand for Paris and {\^I}le-de-France based on open and publicly available data}},
  author={H{\"o}rl, Sebastian and Balac, Milos},
  journal={Transportation Research Part C: Emerging Technologies},
  volume={130},
  pages={103291},
  year={2021},
  publisher={Elsevier}
}

@article{saoPauloEqasim,
  title={{An open data-driven approach for travel demand synthesis: an application to S{\~a}o Paulo}},
  author={Sallard, Aurore and Bala{\'c}, Milo{\v{s}} and H{\"o}rl, Sebastian},
  journal={Regional Studies, Regional Science},
  volume={8},
  number={1},
  pages={371--386},
  year={2021},
  publisher={Taylor \& Francis}
}

@inproceedings{californiaEqasim,
  title={{Synthetic population for the state of California based on open-data: examples of San Francisco Bay area and San Diego County}},
  author={Balac, Milos and H{\"o}rl, Sebastian},
  booktitle={100th Annual Meeting of the Transportation Research Board (TRB)},
  year={2021}
}

@inproceedings{dmc1,
  title={{Pairing discrete mode choice models and agent-based transport simulation with MATSim}},
  author={H{\"o}rl, Sebastian and Bala{\'c}, Milo{\v{s}} and Axhausen, Kay W},
  booktitle={2019 TRB Annual Meeting Online},
  pages={19--02409},
  year={2019},
  organization={Transportation Research Board}
}

@article{dmc2,
  title={{A first look at bridging discrete choice modeling and agent-based microsimulation in MATSim}},
  author={H{\"o}rl, Sebastian and Balac, Milos and Axhausen, Kay W},
  journal={Procedia computer science},
  volume={130},
  pages={900--907},
  year={2018},
  publisher={Elsevier}
}

@article{bottou2018optimization,
  title={{Optimization methods for large-scale machine learning}},
  author={Bottou, L{\'e}on and Curtis, Frank E and Nocedal, Jorge},
  journal={SIAM review},
  volume={60},
  number={2},
  pages={223--311},
  year={2018},
  publisher={SIAM}
}

@article{nakerst2020gradient,
  title={{Gradient descent with momentum---to accelerate or to super-accelerate?}},
  author={Nakerst, Goran and Brennan, John and Haque, Masudul},
  journal={arXiv preprint arXiv:2001.06472},
  year={2020}
}

@article{adam_optimizer,
  title={{Adam: A method for stochastic optimization}},
  author={Kingma, Diederik P and Ba, Jimmy},
  journal={arXiv preprint arXiv:1412.6980},
  year={2014}
}

@article{biogeme,
  title={{A short introduction to Biogeme}},
  author={Bierlaire, Michel},
  journal={Transport and Mobility Laboratory, ENAC, EPFL},
  year={2023}
}

@article{becker2020assessing,
  title={{Assessing the welfare impacts of Shared Mobility and Mobility as a Service (MaaS)}},
  author={Becker, Henrik and Balac, Milos and Ciari, Francesco and Axhausen, Kay W},
  journal={Transportation research part A: policy and practice},
  volume={131},
  pages={228--243},
  year={2020},
  publisher={Elsevier}
}

@article{berlin0,
  title={{The MATSim Open Berlin Scenario: A multimodal agent-based transport simulation scenario based on synthetic demand modeling and open data}},
  author={Ziemke, Dominik and Kaddoura, Ihab and Nagel, Kai},
  journal={Procedia computer science},
  volume={151},
  pages={870--877},
  year={2019},
  publisher={Elsevier}
}

@inproceedings{melbourne,
  title={{Network free flow speed calibration in matsim with google api data: A melbourne study}},
  author={Tiwari, Sapan and Lu, Chengqi and Jafari, Afshin},
  booktitle={Australasian Transport Research Forum},
  year={2024}
}

@article{efficientCalibrationFlotterod,
  title={{Efficient calibration techniques for large-scale traffic simulators}},
  author={Zhang, Chao and Osorio, Carolina and Fl{\"o}tter{\"o}d, Gunnar},
  journal={Transportation Research Part B: Methodological},
  volume={97},
  pages={214--239},
  year={2017},
  publisher={Elsevier}
}

@article{agriesti2023bayesian,
  title={{A Bayesian optimization approach for calibrating large-scale activity-based transport models}},
  author={Agriesti, Serio and Kuzmanovski, Vladimir and Hollm{\'e}n, Jaakko and Roncoli, Claudio and Nahmias-Biran, Bat-Hen},
  journal={IEEE Open Journal of Intelligent Transportation Systems},
  volume={4},
  pages={740--754},
  year={2023},
  publisher={IEEE}
}

@book{dmcbook,
  title={{Discrete Choice Methods with Simulation}},
  author={Train, Kenneth E.},
  year={2009},
  publisher={Cambridge University Press},
  address={Cambridge, UK}
}

@article{horl2021integrating,
  title={{Integrating discrete choice models with MATSim scoring}},
  author={H{\"o}rl, Sebastian},
  journal={Procedia Computer Science},
  volume={184},
  pages={704--711},
  year={2021},
  publisher={Elsevier}
}

@article{horl2021simulation,
  title={{Simulation of price, customer behaviour and system impact for a cost-covering automated taxi system in Zurich}},
  author={H{\"o}rl, Sebastian and Becker, Felix and Axhausen, Kay W},
  journal={Transportation Research Part C: Emerging Technologies},
  volume={123},
  pages={102974},
  year={2021},
  publisher={Elsevier}
}

@inproceedings{spgem,
  title={{Deep learning-based mesoscopic pollutant emissions modeling in road traffic networks}},
  author={Dib, Abdelkader and Sciarretta, Antonio and Balac, Milos},
  booktitle={2024 Forum for Innovative Sustainable Transportation Systems (FISTS)},
  pages={1--7},
  year={2024},
  organization={IEEE}
}

@article{tang2015carbon,
  title={{Carbon emissions trading scheme exploration in China: A multi-agent-based model}},
  author={Tang, Ling and Wu, Jiaqian and Yu, Lean and Bao, Qin},
  journal={Energy Policy},
  volume={81},
  pages={152--169},
  year={2015},
  publisher={Elsevier}
}

@misc{microcensus,
  title        = {{Mikrozensus Mobilität und Verkehr (MZMV)}},
  howpublished = {\url{https://www.bfs.admin.ch/bfs/de/home/statistiken/mobilitaet-verkehr/erhebungen/mzmv.html}},
  year         = {2025},
  note         = {Accessed: 25 August 2025},
  key = {BFS, 2015}
}
\bibliographystyle{elsarticle-harv}



%% \begin{thebibliography}{}


\clearpage

\end{document}